\documentclass[
aps,
prd,
showpacs,
preprint,
tightenlines,
nofootinbib,
floatfix]
{revtex4}
\usepackage{graphicx,
longtable
}

\begin{document}
\title{	Observables sensitive to absolute neutrino masses:
		\\ A reappraisal after WMAP-3y and first MINOS results}
\author{G.L.~Fogli$^1$, E.~Lisi$^1$, A.~Marrone$^1$, A.~Melchiorri$^2$, 
		A.~Palazzo$^{3,1}$, P.~Serra$^2$, J.~Silk$^3$, A.~Slosar$^4$}
\address{
		$^1$~Dipartimento di Fisica and Sezione INFN
		di Bari, Via Amendola 173, 70126, Bari, Italy\smallskip \\ 
		$^2$~Dipartimento di Fisica and Sezione INFN, Universit\`a
		degli Studi di Roma ``La Sapienza'', P.le Aldo Moro 5, 00185, Rome, Italy \smallskip\\
		$^3$~Astrophysics, Denys Wilkinson Building, Keble Road, 
		Oxford, OX1 3RH, United Kingdom \smallskip \\
		$^4$~Faculty of Mathematics and Physics, University of Ljubljana, Slovenia \medskip\medskip}


\begin{abstract}
\medskip
In the light of recent neutrino oscillation and non-oscillation data, 
we revisit the phenomenological constraints applicable to
three observables sensitive to absolute neutrino masses: The
effective neutrino mass in single beta decay $(m_\beta)$; the
effective Majorana neutrino mass in neutrinoless double beta decay
$(m_{\beta\beta})$; and the sum of neutrino masses in cosmology
$(\Sigma)$. In particular, we include the constraints coming from
the first Main Injector Neutrino Oscillation Search (MINOS) data
and from the Wilkinson Microwave Anisotropy Probe (WMAP) three-year (3y) 
data, as well as other relevant cosmological data
and priors. We find that the largest neutrino squared mass 
difference is determined with a  15\% accuracy (at $2\sigma$)
after adding MINOS to world data.
We also find upper bounds on the sum of neutrino masses $\Sigma$
ranging from $\sim2$~eV (WMAP-3y data only) to $\sim0.2$~eV (all cosmological data)
at $2\sigma$, 
in agreement with previous studies. In addition,
we discuss the connection of such bounds with those placed on
the matter power spectrum normalization parameter $\sigma_8$. 
We show how the partial degeneracy 
between $\Sigma$ and $\sigma_8$ in WMAP-3y data is broken 
by adding further cosmological data, and how the
overall preference of such data for relatively high values of $\sigma_8$
pushes the upper bound of $\Sigma$ in the sub-eV range.
Finally, for various combination of data sets, 
we revisit the (in)compatibility between current $\Sigma$ and 
$m_{\beta\beta}$ constraints (and claims), and derive 
quantitative predictions for 
future single and double beta decay experiments.
\end{abstract}
 \maketitle


\section{Introduction}

One of the greatest challenges in current neutrino physics is to establish
the absolute masses $(m_1,m_2,m_3)$ of the three neutrino mass eigenstates 
$(\nu_1,\nu_2,\nu_3)$, for which there is ample experimental
 evidence of mixing
with the three neutrino flavor eigenstates $(\nu_e,\nu_\mu,\nu_\tau)$
through a unitary matrix $U(\theta_{12},\theta_{23},\theta_{13})$, where
$\theta_{ij}$ are the mixing angles \cite{PDG4}.

Atmospheric, solar, reactor, and accelerator
neutrino oscillation experiments
constrain two squared mass differences, $\delta m^2$
and $\Delta m^2$ (with $\delta m^2\ll \Delta m^2$), parametrized as
\begin{equation}
\label{masses}
(m^2_1,\,m^2_2,\,m^2_3)=\frac{m^2_1+m^2_2}{2}+
\left(
-\frac{\delta m^2}{2},\,+\frac{\delta m^2}{2},\,\pm\Delta m^2
\right)\ ,
\end{equation}
where the cases $+\Delta m^2$  and $-\Delta m^2$ distinguish the
so-called normal hierarchy 
(NH) and inverted hierarchy (IH), respectively. The same data
also measure $\theta_{12}$ and $\theta_{23}$, and place upper bounds on 
$\theta_{13}$ (see \cite{PPNP,Viss} for recent reviews).

Non-oscillation neutrino data from single $\beta$ decay, from
neutrinoless double $\beta$ decay ($0\nu2\beta$), and from 
cosmology, add independent constraints on the absolute neutrino masses,
through their sensitivity to the observables \cite{PDG4}
\begin{eqnarray}
\label{mb}
m_\beta &=& \left[c^2_{13}c^2_{12}m^2_1+c^2_{13}s^2_{12}m^2_2+s^2_{13}m^2_3
\right]^\frac{1}{2}\ ,\\
\label{mbb}
m_{\beta\beta} &=&  
\left|
c^2_{13}c^2_{12}m_1+c^2_{13}s^2_{12}m_2e^{i\phi_2}+s^2_{13}m_3
e^{i\phi_3}\right|\ ,\\
\label{Sigma}
\Sigma &=& m_1+m_2+m_3\ ,
\end{eqnarray}
respectively 
(with $c_{ij}=\cos\theta_{ij}$ and $s_{ij}=\sin\theta_{ij}$, while $\phi_{2,3}$
are unknown Majorana phases). 

In a previous work \cite{fogli}, a global analysis of world oscillation 
and non-oscillation data was performed, with the purpose of showing the interplay 
and the (in)compatibility of different data sets in constraining the 
$(m_\beta,m_{\beta\beta},\Sigma)$ parameter space. In this work, we 
revisit such constraints in the light of two recent relevant developments:
the first results from the 
Main Injector Neutrino Oscillation Search (MINOS) \cite{MINOS}
(see Sec.~II), and the three-year (3y)
results  from the Wilkinson
Microwave Anisotropy Probe (WMAP) \cite{WMAP3} (see sec.~IV), which have 
a direct impact on $\Delta m^2$ and $\Sigma$, respectively.
In Sec.~II we show that, after MINOS, the $\Delta m^2$ parameter is globally
determined with an accuracy of
 15\% at $2\sigma$. In Sec.~V 
we discuss the degeneracy between $\Sigma$
and the cosmological parameter $\sigma_8$ (which 
normalizes the matter power spectrum \cite{PDG4,WMAP3}), as well as the breaking of such
degeneracy 
with the addition of further cosmological data, which strenghten
the $2\sigma$ upper bound on $\Sigma$ from $\sim 2$ eV (WMAP 3y only)
to $\sim 0.2$ eV (all cosmological data). Such limits 
on $\Sigma$ are in agreement with other
recent analyses \cite{WMAP3,Seljak:2006bg,goobar,fukugita,raffelt,cirelli,pastor}.
Older $\beta$ and $0\nu2\beta$ results, including 
the $0\nu2\beta$ signal claim of \cite{Kl04}, are  also briefly
reviewed (see Sec.~III). 
In Sec~V we discuss the issues of the (in)compatibility among
the previous data sets  and of their possible 
combinations and constraints in the parameter space $(m_\beta,m_{\beta\beta},\Sigma)$. 
In particular, we show that
the case with ``WMAP 3y only'' still allows a global combination
(at the $2\sigma$ level)
with the $0\nu2\beta$ signal claimed in \cite{Kl04},  
while the addition of other cosmological data excludes this combination 
(at the same or higher significance level). Quantitative implications
for future $\beta$ and $0\nu2\beta$ decay searches are worked out
for various combinations of current data  (Secs.~V and VI). Conclusions
are given in Sec.~VII.


\section{Oscillation parameters after first MINOS results}

The oscillation
parameters $(\delta m^2,\,\theta_{12})$ are essentially determined 
by the solar and KamLAND reactor neutrino data combination, which we take
from \cite{PPNP} (not altered by slightly updated Gallium data
\cite{gavrin}, as we have checked). The corresponding $2\sigma$ ranges  are \cite{PPNP}:
\begin{eqnarray}
\label{limit12a}
\delta m^2 &=& 7.92\,(1\pm 0.09)\times 10^{-5}\mathrm{\ eV}^2\ ,\\
\label{limit12b}
\sin^2\theta_{12}&=&0.314\,(1^{+0.18}_{-0.15})\ .
\end{eqnarray}

The parameters 
($\Delta m^2,\sin^2\theta_{23}$) are  dominated
by atmospheric and accelerator $\nu$ data,
within the CHOOZ \cite{CHOOZ} bounds on $\sin^2\theta_{13}$ 
(see \cite{PPNP} for details). 
After the review 
\cite{PPNP}, oscillation results have been finalized for 
the Super-Kamiokande (SK) atmospheric $\nu$ experiment 
\cite{SKatm} and for
the KEK-to-Kamioka (K2K) accelerator $\nu$ experiment \cite{K2K}. 
The SK+K2K analysis in \cite{PPNP}, however, contains most of the 
final SK and K2K statistics, and is not updated here.

New accelerator data in the $\nu_\mu\to\nu_\mu$ disappearance channel
have recently been released by the MINOS experiment \cite{MINOS}.
The MINOS data significantly help 
to constrain the $\Delta m^2$ parameter \cite{MINOS}, while they are less
sensitive to $\theta_{23}$ (as compared with atmospheric data \cite{SKatm}),
and are at present basically insensitive to $\sin^2\theta_{13}$, 
although some sensitivity will be gained through future
searches in the  $\nu_\mu\to\nu_e$ appearance channel. They are also
insensitive to the parameters $(\delta m^2,\,\theta_{12})$, whose
main effect (at $\theta_{13}\simeq 0$) is to change the oscillation phase 
by a tiny fractional amount $0.5\, \delta m^2 \cos2\theta_{12}/\Delta m^2\simeq 0.6\times 10^{-2}$
(see, e.g., \cite{PPNP}), which is negligible within current uncertainties. 
Therefore, the usual
two-family approximation---also
adopted in the official MINOS analysis \cite{PPNP}---is 
currently appropriate to study MINOS data.

In the absence of a detailed description
of the procedure used by the MINOS collaboration
for the data analysis, we provisionally include their constraints on 
the parameters $(\Delta m^2,\,\sin^2\theta_{23})$ through
an empirical parametrization of the $\chi^2$ statistical function,
\begin{equation}\label{chiminos}
\chi^2_\mathrm{MINOS}=\left(\frac{x-1}{0.134}\right)^2+
\left(\frac{y\cdot x^{0.76}-2.74}{0.27}\right)^2\ ,
\end{equation}
where $x=\sin^2(2\theta_{23})$ and $y=\Delta m^2/(10^{-3}\ \mathrm{eV}^2)$. This 
parametrization accurately reproduces the official MINOS bounds
in the plane ($\Delta m^2,\sin^2 2\theta_{23}$) \cite{MINOS}
(at least in the relevant region at relatively large mixing angles, 
$\sin^2 \theta_{23}\in [0.3,0.7]$). A more proper analysis
will be performed when further experimental and statistical
details will be made public by MINOS.

By adding the above $\chi^2$ function in the global analysis 
of neutrino oscillation data performed in \cite{PPNP}, we obtain the following
$2\sigma$ allowed ranges:
\begin{eqnarray}
\label{limit23a}
\Delta m^2 &=& 2.6\,(1^{+0.14}_{-0.15})\times 10^{-3}\mathrm{\ eV}^2\ ,\\
\label{limit23b}
\sin^2\theta_{23}&=&0.45\,(1^{+0.35}_{-0.20})\ ,
\end{eqnarray}
which noticeably improve the previous bounds in \cite{PPNP}. In particular,
the $2\sigma$ error on $\Delta m^2$ is reduced from $\sim 24\%$ \cite{PPNP}
to $\sim 15\%$ after the inclusion of MINOS data.

The error reduction for $\Delta m^2$ also reduces the spread
of the (dominant) CHOOZ limits on $\sin^2\theta_{13}$ \cite{CHOOZ}, which are 
$\Delta m^2$-dependent. Therefore, MINOS indirectly provides a slight
improvement of the previous bounds on $\sin^2\theta_{13}$ \cite{PPNP},
which are now updated at $2\sigma$  as:
\begin{equation}
\label{limit13}
\sin^2\theta_{13}=(0.8^{+2.3}_{-0.8})\times 10^{-2}\ .
\end{equation}
Equations~(\ref{limit12a}), (\ref{limit12b}), (\ref{limit23a}), (\ref{limit23b}) 
and (\ref{limit13}) represent our up-to-date evaluation of the neutrino
oscillation parameters (at 95\% C.L.).

\begin{figure}
\vspace*{-0cm}\hspace*{-.2cm}
\includegraphics[scale=0.95]{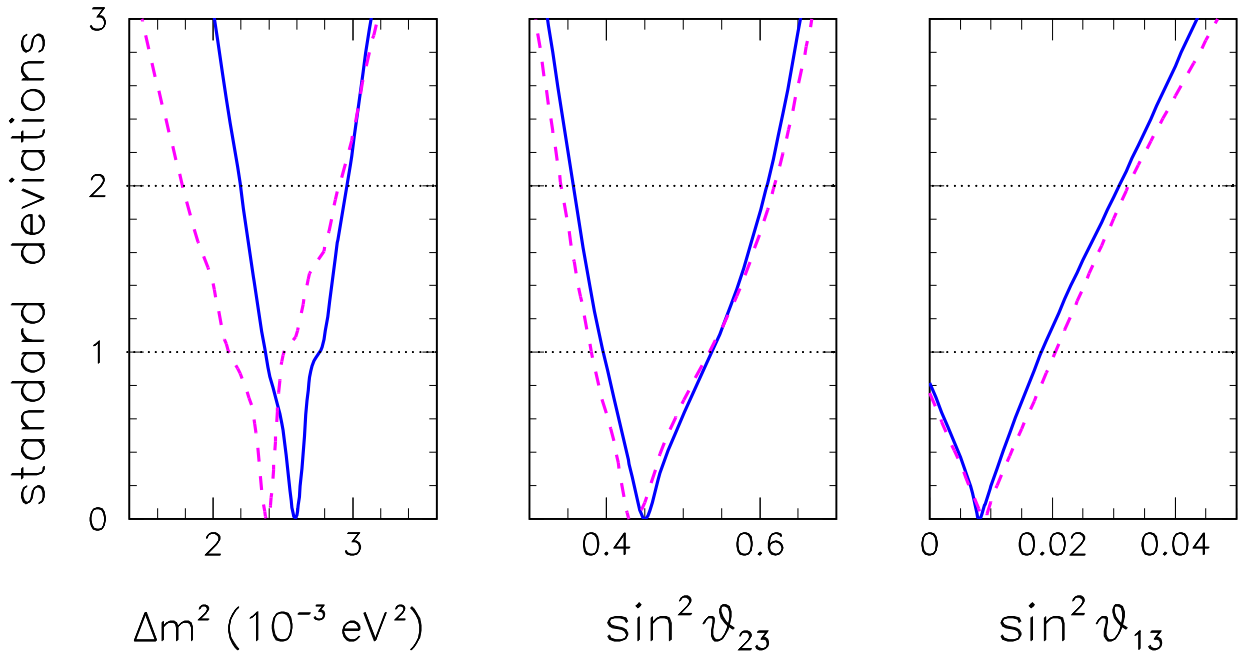}
\caption{\label{fig1} 
Constraints placed by neutrino oscillation data on the parameters  
($\Delta m^2,\,\sin^2\theta_{23},\,\sin^2\theta_{13}$), which
are affected by the inclusion of the first MINOS results. The results are shown
in terms of standard deviations from the
best fit. Solid (dashed) lines refers to all neutrino oscillation data
with (without) MINOS. 
}
\end{figure}

Figure~\ref{fig1} shows the comparison between the limits obtained on
$(\Delta m^2,\,\sin^2\theta_{23},\,\sin^2\theta_{13})$
before MINOS (dashed, from \cite{PPNP}) and after MINOS (solid, this work), in terms
of standard deviations from the best fit (i.e., in terms of $\sqrt{\Delta\chi^2}$,
where $\Delta\chi^2$ is the fitting function to all oscillation data 
\cite{PPNP}).
The impact of MINOS on the $\Delta m^2$ parameter is rather 
significant, and the related uncertainties appear now to scale 
almost linearly and symmetrically. The analogous figure (not shown) for
the $(\delta m^2,\,\sin^2\theta_{12})$ parameters---which are not
affected by MINOS data---can be found in \cite{PPNP}.

\section{Input from single and double beta decay searches}

Single $\beta$-decay experiments probe the effective electron neutrino mass
parameter $m_\beta$, at least in first approximation. The most stringent
constraints on $m_\beta$ are placed by the Mainz and Troitsk
experiments \cite{Eite}, the latter being affected by still unexplained time-dependent
systematics. We assume that both the Mainz and Troitsk results can be
taken at face value and combined as in \cite{PPNP,fogli}, 
obtaining at $2\sigma$:
\begin{equation}
m_\beta < 1.8 \mathrm{\ eV}\ .
\end{equation}
If only the Mainz results were used, this limit would be slightly relaxed 
($m_\beta<2.2$ eV). In any case, the above bound is relatively weak
and has almost no impact in the following analyses, except for
one scenario (see Case $1^+$ in Sec.~V~A).

Double beta decay searches with no final-state neutrinos
have not reported positive signals so far \cite{Elli},
except in the most sensitive ($^{76}$Ge) detector to date (Heidelberg-Moscow experiment), where
part of the collaboration has claimed a signal at $>4\sigma$ level 
\cite{Kl04}, recently
promoted to $>6\sigma$ level by a pulse-shape analysis \cite{Kl06}. 
This claim is still considered as controversial \cite{Elli} 
and we shall discuss its implications with due care
in the following sections.
In any case, any claim or limit on the $0\nu2\beta$ decay half-life ($T^{0\nu}_{1/2}$) 
in a candidate nucleus constrains
the effective Majorana mass ($m_{\beta\beta}$) through the relation
\begin{equation}
\label{nonlog}
m^2_{\beta\beta}=\frac{m^2_e}{C_{mm}\,T^{0\nu}_{1/2}}\ ,
\end{equation}
in the assumption that the $0\nu2\beta$ process proceeds {\em only\/}
through light Majorana neutrinos (and not through new interactions or particles) \cite{Elli}. 
The relevant nuclear physics is included in the
matrix element $C_{mm}$, which must be theoretically calculated.

Since the above relation is non-linear, and since all the quantities involved
(except the electron mass $m_e$) are subject to large uncertainties, we prefer
to linearize Eq.~(\ref{nonlog}) and to deal with more ``tractable'' 
uncertainties by using logarithms as in \cite{fogli}:
\begin{equation}
\label{logs}
2\log_{10}\left(\frac{m_{\beta\beta}}{\mathrm{eV}}\right)=
2\log_{10}\left(\frac{m_e}{\mathrm{eV}}\right)-
\log_{10}\left(\frac{C_{mm}}{\mathrm{yr}^{-1}}\right)-
\log_{10}\left(\frac{T^{0\nu}_{1/2}}{\mathrm{yr}}\right)
\ .
\end{equation}
In the following, it is understood that linear error propagation is applied to the above ``logs'' 
(e.g., $\log C_{mm}$) rather than to the exponentiated quantities (e.g., $C_{mm}$).

\begin{table}[t]
\caption{\label{tableNME} Our summary of the estimated $0\nu2\beta$ 
nuclear matrix elements $C_{mm}$ and their $\pm2\sigma$ errors
(in logarithmic scale) for seven relevant
nuclei, as derived from Ref.~\cite{faessler}. See the text for details.}
\centering
\resizebox{\textwidth}{!}{
\begin{ruledtabular}
\begin{tabular}{cc}
Nucleus		& $\log_{10}(C_{mm}/\mathrm{yr}^{-1})\pm 2\sigma$ \\[4pt]
\hline
$^{76}$Ge	& $-13.36\pm0.10$		\\
$^{82}$Se	& $-12.83\pm0.14$		\\
$^{100}$Mo	& $-13.13\pm0.20$		\\
$^{116}$Cd	& $-12.96\pm0.24$		\\
$^{128}$Te	& $-14.32\pm0.29$		\\
$^{130}$Te	& $-12.98\pm0.30$		\\
$^{136}$Xe	& $-13.41\pm0.56$		
\end{tabular}
\end{ruledtabular}
}\vspace*{0.8cm}
\end{table}


In our analysis, we take the theoretical input 
for $\log C_{mm}$ (central values and errors) 
from the quasi-random phase approximation (QRPA)
calculations in
Ref.~\cite{faessler}, where it has been shown that the nuclear 
uncertainties can be significantly constrained (and reduced) by
requiring consistency with independent $2\nu2\beta$ decay data
(when available). More precisely, for a given nucleus,
we take the logarithms of the minimum and
maximum values of $C_{mm}$ from the $\pm 1\sigma$ ranges
reported in the last column of Table~1 in \cite{faessler}; then, by taking
half the sum and half the difference of these extremal values
we get, respectively, 
our default central value
and $1\sigma$ error for $\log C_{mm}$ ($2\sigma$ errors are just doubled). 
Table~\ref{tableNME} shows a summary of the $\log C_{mm}$ values and 
their $2\sigma$ errors derived in this way, and used hereafter. 

With the theoretical input for $C_{mm}({}^{76}\mathrm{Ge})$ from Table~\ref{tableNME}, 
the $0\nu2\beta$ claim of \cite{Kl04} is transformed in the following  $2\sigma$ range for 
$m_{\beta\beta}$:
\begin{equation}
\log_{10}(m_{\beta\beta}/\mathrm{eV})= -0.23\pm 0.14\ \
(0\nu2\beta \mathrm{\ claim\  accepted})
\label{logmbb2}\ ,
\end{equation}
i.e.,
$0.43<m_{\beta\beta}<0.81$ (at $2\sigma$, in eV). See also \cite{fogli}
for our previous estimated range (with more conservative theoretical uncertainties).
As in \cite{fogli}, we  consider the
possibility that $T^{0\nu}_{1/2}=\infty$ is allowed (i.e., that the claimed 
$0\nu2\beta$ signal is incorrect), in which case the experimental lower bound on
$m_{\beta\beta}$ disappears, and only the upper bound at $2\sigma$
remains:
\begin{equation}
\log_{10}(m_{\beta\beta}/\mathrm{eV})=-0.23^{+0.14}_{-\infty}\
(0\nu2\beta \mathrm{\ claim\ not\ accepted})\ . \label{bbinput2}
\end{equation}

\vspace*{-0.4cm}


\section{Input from cosmological data}

The neutrino contribution to the overall energy density of the
universe can play a relevant role in large scale structure formation
and leave key signatures in several cosmological data sets. More
specifically, neutrinos suppress the growth of fluctuations on
scales below the horizon when they become non relativistic. 
Massive neutrinos with $\Sigma=m_1+m_2+m_3$ in the (sub)eV range
would then produce a
significant suppression in the clustering on small cosmological
scales (see \cite{pastor} for a recent review).

The method that we adopt to derive bounds on $\Sigma$
 is based on the publicly available Markov Chain Monte Carlo (MCMC)
package \texttt{cosmomc} \cite{Lewis:2002ah}. We sample the following
 set of cosmological parameters, adopting flat priors on them:
the physical baryon, cold dark matter, and massive neutrino densities ($\omega_b=\Omega_bh^2$,
$\omega_c=\Omega_ch^2$ and $\Omega_{\nu}h^2$, respectively),
 the ratio of the sound horizon to the angular diameter
distance at decoupling ($\theta_s$), the scalar spectral index, 
the overall normalization of the spectrum $A$ at wavenumber
$k=0.05$ Mpc$^{-1}$ and, finally, the optical
depth to reionization, $\tau$. Furthermore, we consider purely adiabatic
initial conditions and we impose flatness.

 From a technical viewpoint,
we include the WMAP 3y data (temperature and polarization)
\cite{WMAP3,wmap3temp}
with the routine for computing the likelihood supplied by the WMAP team 
\cite{LAMBDA}. We marginalize over the amplitude
of the Sunyaev-Zel'dovich signal, but the effect is small: including
or excluding
such correction change our best fit values for single
parameters by less than $2\%$ and always well inside the $1 \sigma$ confidence
level. We treat beam errors with the highest possible accuracy (see
\cite{wmap3temp}, Appendix A.2), using full off-diagonal temperature
covariance matrix, Gaussian plus lognormal likelihood, and fixed fiducial
$C_{\ell}$'s. The MCMC convergence diagnostics is done throught the so-called
Gelman and Rubin  
``variance of chain mean''$/$``mean of chain variances'' $R$
ratio statistic for each variable. Our final constraints over one 
parameter ($\Sigma$) or two parameters ($\Sigma,\sigma_8$) are obtained
after marginalization over all the other ``nuisance'' parameters, again using
the programs included in the \texttt{cosmomc} package. In addition to 
Cosmic Microwave Background (CMB)
data from WMAP, we also include other relevant data semples, 
according to the
following numbered cases:

\begin{description}

\item[1)] {\bf WMAP only:} 
Only temperature, cross polarization and
  polarization WMAP 3y data are considered, plus a top-hat age prior 
$10 \mathrm{\ Gyr} <  t_0 < 20 \mathrm{\ Gyr}$.

\item[2)] {\bf WMAP+SDSS:}
We combine the WMAP data with the the real-space power spectrum of
galaxies from the Sloan Digital Sky Survey (SDSS) 
\cite{2004ApJ...606..702T}. We restrict the analysis to a range of scales over which the
fluctuations are assumed to be in the linear regime (technically, $k < 0.2
h^{-1}$~Mpc) and we marginalize over a  bias $b$ considered as an
additional nuisance parameter.

\item[3)] {\bf  WMAP+SDSS+SN$_\mathrm{Riess}$+HST+BBN:} 
We combine the data considered in the previous case with 
Hubble Space Telescope (HST)
 measurement of the Hubble parameter $H_0 = 100h \text{ km s}^{-1}\, \text{Mpc}^{-1}$
\cite{hst}, a Big Bang Nucleosynthesis prior 
$\Omega_bh^2=0.020\pm0.002$, and we finally incorporate the constraints
obtained from the supernova (SN-Ia) luminosity measurements of \cite{riess} by using 
the so-called GOLD data set.

\item[4)]{\bf  CMB+LSS+SN$_\mathrm{Astier}$:} Here we include WMAP data
and also consider the small-scale CMB measurements of the CBI 
\cite{2004ApJ...609..498R}, VSA \cite{2004MNRAS.353..732D}, ACBAR
\cite{2002AAS...20114004K} and BOOMERANG-2k2
\cite{2005astro.ph..7503M} experiments.  In addition to
the CMB data, we include the large scale structure (LSS)
constraints on the real-space power
spectrum of galaxies from the SLOAN galaxy redshift survey (SDSS) 
\cite{2004ApJ...606..702T} and 2dF survey \cite{2005MNRAS.362..505C}, 
 as well as the Supernovae Legacy Survey data from \cite{2006A&A...447...31A}.
 
\item[5)]{\bf  CMB+LSS+SN$_\mathrm{Astier}$+BAO:} We include data as in the previous case plus
the constraints from the Baryonic Acoustic Oscillations (BAO) detected in
the Luminous Red Galaxies sample of
the SDSS \cite{2005ApJ...633..560E}.

\item[6)]{\bf  CMB+SDSS+SN$_\mathrm{Astier}$+Lyman-$\alpha$:} We include measurements of
the small scale primordial spectrum from 
Lyman-alpha (Ly-$\alpha$) forest clouds \cite{McDonald:2004eu,McDonald:2004xn}
but we exclude BAO constraints. The details of the analysis are the
same as those in \cite{Seljak:2006bg,Dodelson:2005tp}.

\item[7)]{\bf  CMB+SDSS+SN$_\mathrm{Astier}$+BAO+Lyman-$\alpha$:} We 
add the  
BAO measurements to the previous dataset. Again, see
 \cite{Seljak:2006bg,Dodelson:2005tp} for more details.

\end{description}

\begin{figure}[t]
\vspace*{-0cm}\hspace*{-0.5cm}
\includegraphics[scale=0.6]{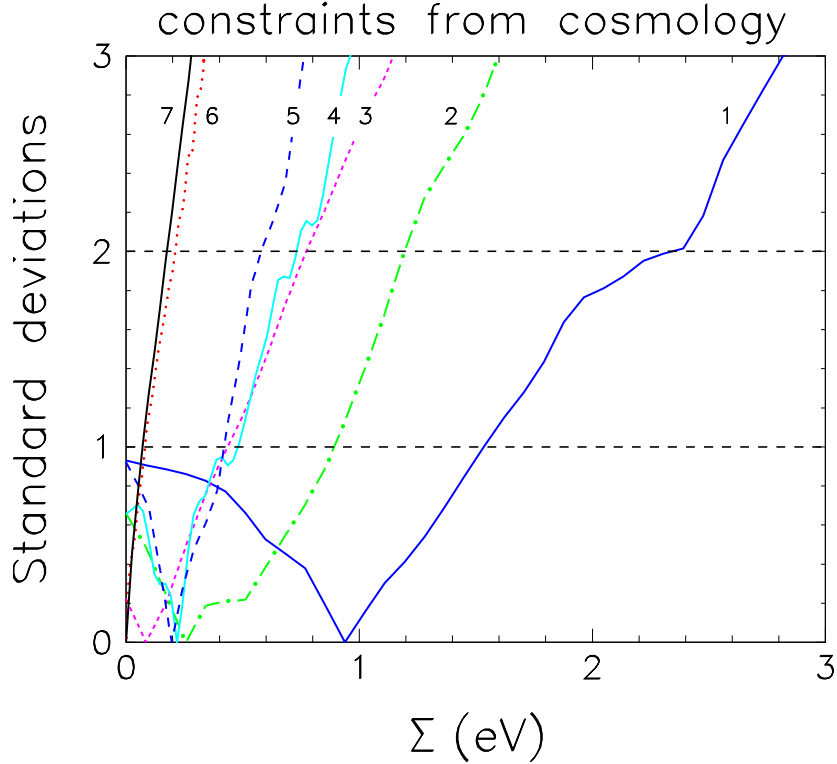}
\vspace*{-0.2cm}
\caption{\label{fig2} 
Constraints placed by different cosmological data sets (1,...,7) on the
sum of neutrino masses $\Sigma$, in terms of standard deviations from the
best fit in each case.
}
\end{figure}

The above cases provide a sufficiently rich list of cosmological
data combinations, with increasingly strong constraints on $\Sigma$.
In particular, Fig.~\ref{fig2} shows the constraints on 
$\Sigma$ for each case of our analysis, in terms of 
standard deviations from the best fit of $\Sigma$. 
None of these curves shows evidence for  neutrino mass
at $>1\sigma$, indicating that current cosmological
data can only set upper bounds on $\Sigma$. The bounds  tend
to scale linearly for the richest data sets (e.g., for the cases 5, 6, and 7).

\begin{table}[t]
\caption{\label{tableCASES} Input cosmological data sets for seven representative
cases considered in this work, together with their $2\sigma$ (95\% C.L.) constraints 
on the sum of neutrino masses $\Sigma$.}
\centering
\resizebox{\textwidth}{!}{
\begin{ruledtabular}
\begin{tabular}{lll}
Case		& Cosmological data set							& $\Sigma$ bound ($2\sigma$)\\[4pt]
\hline
1 & WMAP 												& $<2.3$ eV \\
2 & 	WMAP + SDSS											& $<1.2$ eV \\
3 & 	WMAP + SDSS + SN$_\mathrm{Riess}$ + HST + BBN			& $<0.78$ eV \\
4 & 	CMB	+ LSS + SN$_\mathrm{Astier}$ 					& $<0.75$ eV \\
5 & 	CMB	+ LSS + SN$_\mathrm{Astier}$ + BAO				& $<0.58$ eV \\
6 & 	CMB	+ LSS + SN$_\mathrm{Astier}$ + Ly-$\alpha$		& $<0.21$ eV \\
7 & 	CMB	+ LSS + SN$_\mathrm{Astier}$ + BAO + Ly-$\alpha$	& $<0.17$ eV 
\end{tabular}
\end{ruledtabular}
}\vspace*{1.2cm}
\end{table}

Table~II summarizes the bounds on $\Sigma$ derived from our analysis
in numerical form (at the $2\sigma$ level).
Such bounds are in agreement with
previous results in similar recent analyses which include
WMAP and other data, whenever a comparison is
possible \cite{WMAP3,Seljak:2006bg,goobar,fukugita,raffelt,cirelli,pastor}, and we can 
derive the following conclusions:

\begin{itemize}

\item As already shown in \cite{WMAP3} and
\cite{fukugita}, the WMAP data alone,
in the framework of cosmological models we are considering,
are able to constrain $\Sigma \lesssim 2$~eV at $95 \%$ C.L..
This limit, which relyes on a single high-quality dataset,
should be considered as the most conservative.

\item Inclusions of galaxy clustering and SN-Ia data can,
as already pointed out in the literature (see e.g.\ \cite{fogli,pastor} and
references therein), further constrain the
results. The datasets used in the cases $2$, $3$, $4$ and $5$ contain
different galaxy clustering and supernovae data and thus test the
impact of possible different systematics. Such cases 
provide, respectively, constraints of $\Sigma <1.2$~eV, $\Sigma < 0.78$~eV,
$\Sigma < 0.75$~eV and $\Sigma < 0.58$~eV at $95 \%$~C.L., respectively. 
These results are in reasonable agreement with the findings of, e.g., \cite{WMAP3}
but are slightly weaker than those presented in \cite{goobar} (case $5$ in particular), 
probably as a consequence of the different
data analysis method. Just to mention few differences, in 
\cite{goobar} the likelihood analysis 
is based on a database of models while we adopt
Markov Chains. Moreover \cite{goobar} includes variations
in the equations of state parameter $w$ and running of the spectral
index, while here we consider only models with $w=-1$ and no running. 
Non-standard models with a large negative running of the spectral index or with $w>-1$ 
prefer smaller neutrino masses when compared with observations.
Finally, \cite{goobar} uses the same bias parameter
for the SLOAN and 2dF surveys while the two may be different.

\item Including SDSS Ly-$\alpha$ data in cases $6$ and $7$ (as in
\cite{Seljak:2006bg}) greatly improves
the constraints on $\Sigma$ up to $\Sigma < 0.21$eV 
and $\Sigma < 0.17$eV ($95 \%$ C.L.). The latter is stronger
than the one reported in \cite{goobar}, since we are using the updated
SDSS Ly-$\alpha$ dataset of \cite{Seljak:2006bg}.
This constraint has important
consequences for our analyses, especially when compared with the
$m_{\beta \beta}$ claim. However, we remark that the bounds
on the linear density fluctuations obtained from the
SDSS Ly-$\alpha$ dataset are
derived from measurement of the Ly-$\alpha$ flux power spectrum
$P_F(k)$ after a long inversion process, which involves numerical
simulations and marginalization over the several parameters of the
Ly-$\alpha$ model. The strongest limit in case 7
should therefore be considered as the less conservative.
\end{itemize}

As we can see, cosmology seems to provide the best constraints
available on absolute neutrino masses. However, sub-eV constraints can
be placed only when we consider a combination of
multiple datasets. It is therefore important to check the
degree of compatibility between the datasets. 
A possible way to make such check and to better understand, at the
same time, the data preference for small values of $\Sigma$, is  
to consider a joint analysis of $\Sigma$ and of the
so-called $\sigma_8$ parameter, which represents 
the expected linear root mean square (rms) amplitude of matter fluctuations
in spheres of radius $R=8h^{-1}$~Mpc. 

Let us briefly remind that the linear (rms) mass fluctuations in spheres
of radius $R$ are usually expressed throught their power spectrum
$P(k)$ in Fourier space (see e.g.\ \cite{bondszalay}):
\begin{equation}
 \sigma^2(R)=\int_0^\infty {dk\over k}\, 4\pi k^3 
	P(k) W^2(kR)\ ,
\label{sigma}
\end{equation}
 where $W(x)=3(\sin x - x \cos x)/x^3$ is the top-hat window function,
and the mass enclosed in the sphere is $M=4\pi\rho_0 R^3/3$, with
$\rho_0$ denoting the background mass density of the universe.
The matter power spectrum $P(k)$ is fully determined once the
cosmological model and the corresponding parameters are defined.
The effect of massive neutrinos is to reduce the amplitude
of the power spectrum $P(k)$ on free streaming scales 
and, therefore, to reduce the value of $\sigma_8$. 
The neutrino
free-streaming process introduces indeed an additional
length scale in the power spectrum  related to
the median neutrino Fermi-Dirac speed $v_{\rm med}$
by  $k_{fs}^2=4\pi G\rho/v^2_{\rm med}\propto \Sigma h$.  
For $k<k_{fs}$, the density perturbation in the neutrinos grows in the 
matter-dominated era while decays for $k>k_{fs}$. 
It is possible to show (see e.g. \cite{tegmax}) that for
 wavenumbers smaller than $k_{fs}$ the power specrum  is damped by a factor
\begin{equation}
P(k;\Sigma)/P(k;0) \approx e^{-0.087 (\Sigma/{\Omega_mh^2})}\ ,
\end{equation}
where $\Omega_m$ is the total matter density.
Therefore, in general, we expect that the larger $\Sigma$ 
the smaller $\sigma_8$ (or vice versa), namely, that these 
two parameters are partly degenerate, with negative correlations. 
If $\sigma_8$ is determined
in some way, the degeneracy is ``broken'' and $\Sigma$ can be better
constrained.

The value of $\sigma_8$ can be derived either  in an {\it indirect\/} way,
i.e., by first determining the cosmological model and by then
considering the possible values of the allowed $P(k)$, or 
in a more {\it direct\/} way, i.e.,  by measuring the amount of galaxy
clustering on $\sim 8$ Mpc scales.
Since CMB observations provide an indirect determination of
$\sigma_8$, it is important to check if this determination is
compatible with the other, more direct, measurements made throught 
galaxy clustering. An incompatibility between the two datasets,
with, for example, a low value preferred by CMB data {\em vs\/} a
high value preferred by
clustering data, would lead to a formally very strong 
(but practically unreliable)
constraint on  $\Sigma$ (see also \cite{WMAP3}).

\begin{figure}
\vspace*{-0cm}\hspace*{-0.5cm}
\includegraphics[scale=1.02]{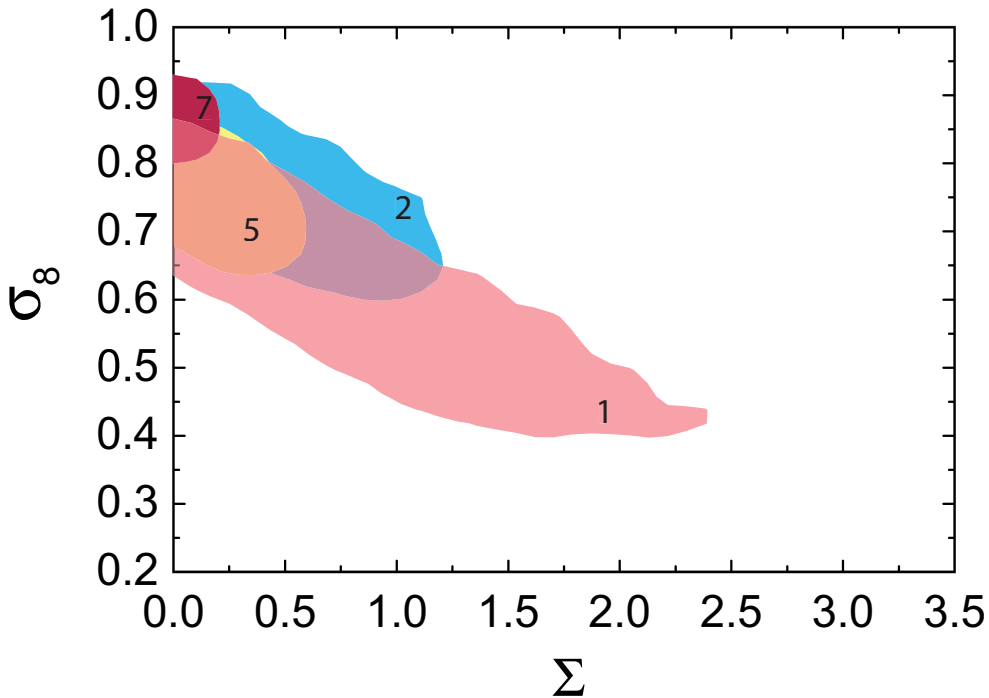}
\caption{\label{fig3} 
Joint $2\sigma$ constraints on the $(\sigma_8,\,\Sigma)$ parameters 
(95\% C.L.\ for $N_\mathrm{DF}=1$)  derived from four representative cosmological data
sets (1, 2, 5, and 7, as listed in Table~II).
}
\end{figure}

Fig.~\ref{fig3} shows the results of our joint analysis of the
$(\sigma_8,\,\Sigma)$ parameters, in terms of  $2\sigma$ contours 
($\Delta\chi^2=4$ for one degree of frreedom) for four representative cosmological data
sets (1, 2, 5, and 7), chosen among
the seven sets studied in this work (see Table~II). In case~1
(WMAP only), the degeneracy and anticorrelation between the two
parameters is evident. The addition of
galaxy clustering and Ly-$\alpha$ tend to break the degeneracy
by selecting increasingly smaller ranges for $\sigma_8$. It turns
out that such additional data tend to prefer $\sigma_8$ in the
``higher part'' of the range allowed by WMAP only, so that the
upper bound on $\Sigma$ is pushed to lower values. Should future
data hypothetically invert such trend (i.e., the current preference
for ``high'' values of $\sigma_8$), the upper bounds
on $\Sigma$ would be somewhat relaxed. In any case, the different
combinations of current cosmological  data in Fig.~3 appear to be
in relatively good agreement with each other, with significant
overlap of the different $2\sigma$ regions---a reassuring 
consistency check.


\section{Global analyses: Results for various input data sets}

In this section we discuss the global analysis of all neutrino oscillation and
non-oscillation data, with particular attention to the (in)compatibility 
and combination (at $2\sigma$ level) 
of the various cosmological data sets (numbered as $1,\, 2,\dots 7$
in Table~\ref{tableCASES}) with the $0\nu2\beta$ signal claim of \cite{Kl04}.
First we examine the most conservative case 
where a combination is possible (case $1^+$ in Sec.~V~A), 
then we discuss the worst case where the combination is forbidden 
at $\gg 2\sigma$ (case 7 in Sec.~V~B),
and finally we make an overview of the results for all possible 
intermediate cases 
considered in this work (Sec.~V~C).

\begin{figure}[t]
\vspace*{-0cm}\hspace*{-.5cm}
\includegraphics[scale=0.73]{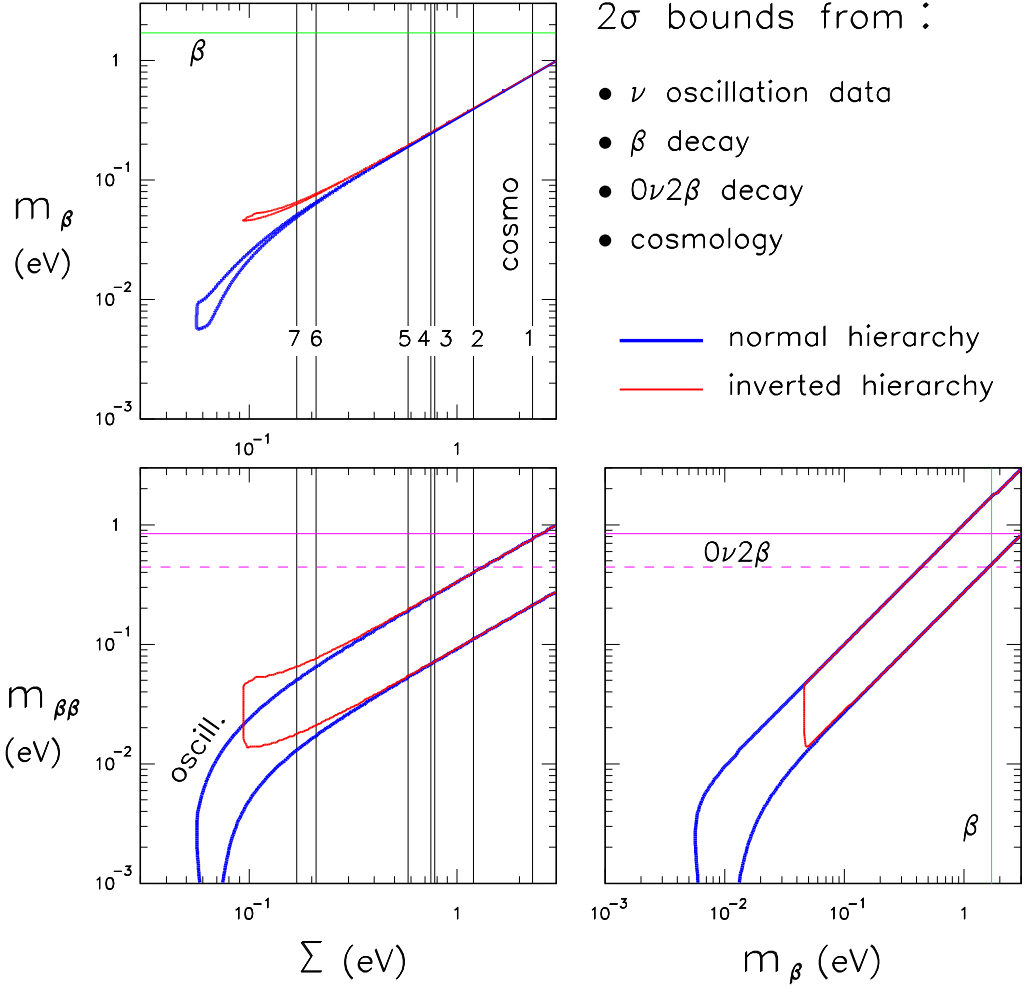}
\caption{\label{fig4} 
Superposition of $2\sigma$ constraints (95\% C.L.\ for $N_\mathrm{DF}$=1) 
placed by $\beta$, $0\nu2\beta$,
oscillation, and cosmological neutrino data in the three 2-dimensional
projections of the $(m_\beta,\,m_{\beta\beta},\,\Sigma)$ parameter space.
Cosmological constraints are labelled as in Table~\ref{tableCASES} for the seven input data
sets. The $0\nu2\beta$ lower limit on $m_{\beta\beta}$ from the claim in \protect\cite{Kl04} is
indicated as a horizontal dashed line.}
\end{figure}

The results discussed in more detail in the following sections
can be qualitatively understood through 
Fig.~\ref{fig4}, which---following the previous work \cite{fogli}---shows the 
three orthogonal projections of the 
regions separately
allowed at $2\sigma$ level in the $(m_{\beta},\,m_{\beta\beta},\,\Sigma)$
parameter space by: (a) neutrino oscillation data (slanted bands for normal
and inverted hierarchy); (b) the seven cosmological data sets (numbered lines
with ``cosmo'' label); (c) single $\beta$ decay searches (lines with $\beta$ label);
and (d) $0\nu2\beta$ decay limits (lines with $0\nu2\beta$ label). 
In the
latter case, the lower limit from the claim in \cite{Kl04} is shown as a dashed line
to remind that, formally, it may be accepted or not. In Fig.~\ref{fig4} the
various bounds are simply superposed, while real combinations of data are
performed in the following sections. Note that the global
combinations can alter the bounds from separate data sets: e.g.,
 the global upper bounds on $\Sigma$ may be
slightly different from those placed by cosmological data only,
while lower bounds on $\Sigma$ (not placed at all by current cosmological data)
arise when the oscillation data (i.e., the evidence for nonzero neutrino mass)
 is included.

\begin{figure}[t]
\vspace*{-0cm}\hspace*{-.5cm}
\includegraphics[scale=0.7]{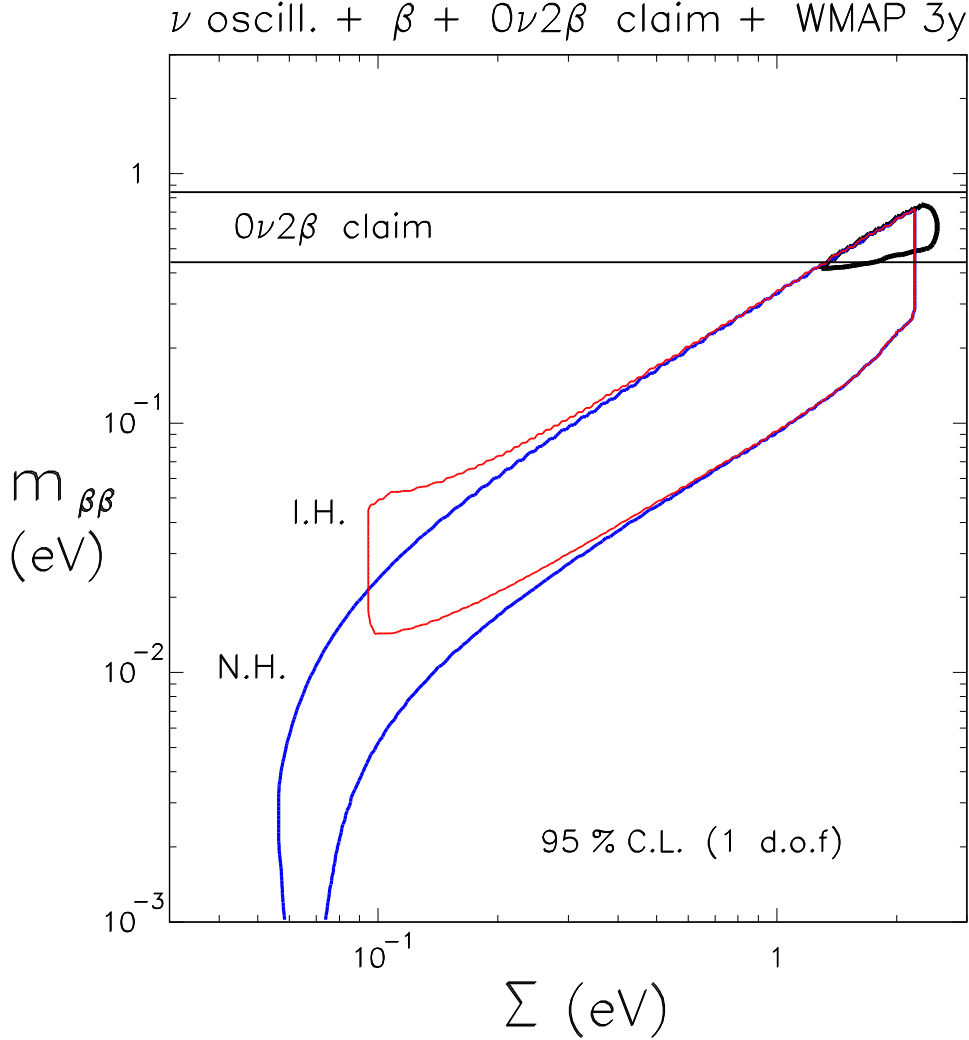}
\vspace*{2mm}
\caption{\label{fig5} 
Comparison (at $2\sigma$) between the regions preferred by
the $0\nu2\beta$ signal claim (horizontal band) and by
 all $\nu$ oscillation data plus $\beta$ and WMAP-3y data (slanted
bands) for normal (NH) and inverted (IH) hierarchy, in the
plane ($m_{\beta\beta},\Sigma$). The 
combination of all such data (thick slanted ``wedge'' in the upper
right corner of the plot) corresponds
to ``Case $1^+$'' in the text.
}
\end{figure}

\subsection{Case with WMAP-3y data and the $0\nu2\beta$ claim}

The most conservative choice for the cosmological input is to rely only
upon WMAP 3y data (case 1 in Table~\ref{tableCASES}). In this case, the upper bound 
on $\Sigma$ is relatively weak ($\sim 2$ eV), and the ``slanted band'' allowed by the 
combination of oscillation and cosmological data in Fig.~4 can reach the horizontal
 band allowed by the claimed $0\nu2\beta$ signal \cite{Kl04}, and
 a global combination of all data becomes possible. Figure~5 shows in more
detail this situation in the relevant plane $(m_{\beta\beta},\Sigma)$, 
where the global combination (from a full $\chi^2$ analysis) yields 
a ``wedge shaped'' allowed region around  $\Sigma\sim 2$~eV and 
$m_{\beta\beta}\sim\mathrm{few}\times 10^{-1}$ eV.
This case will be labelled ``$1^+$'' in the following (meaning:
cosmological data set 1, plus $0\nu2\beta$ claim).

\begin{figure}[t]
\vspace*{-0cm}\hspace*{-.5cm}
\includegraphics[scale=0.7]{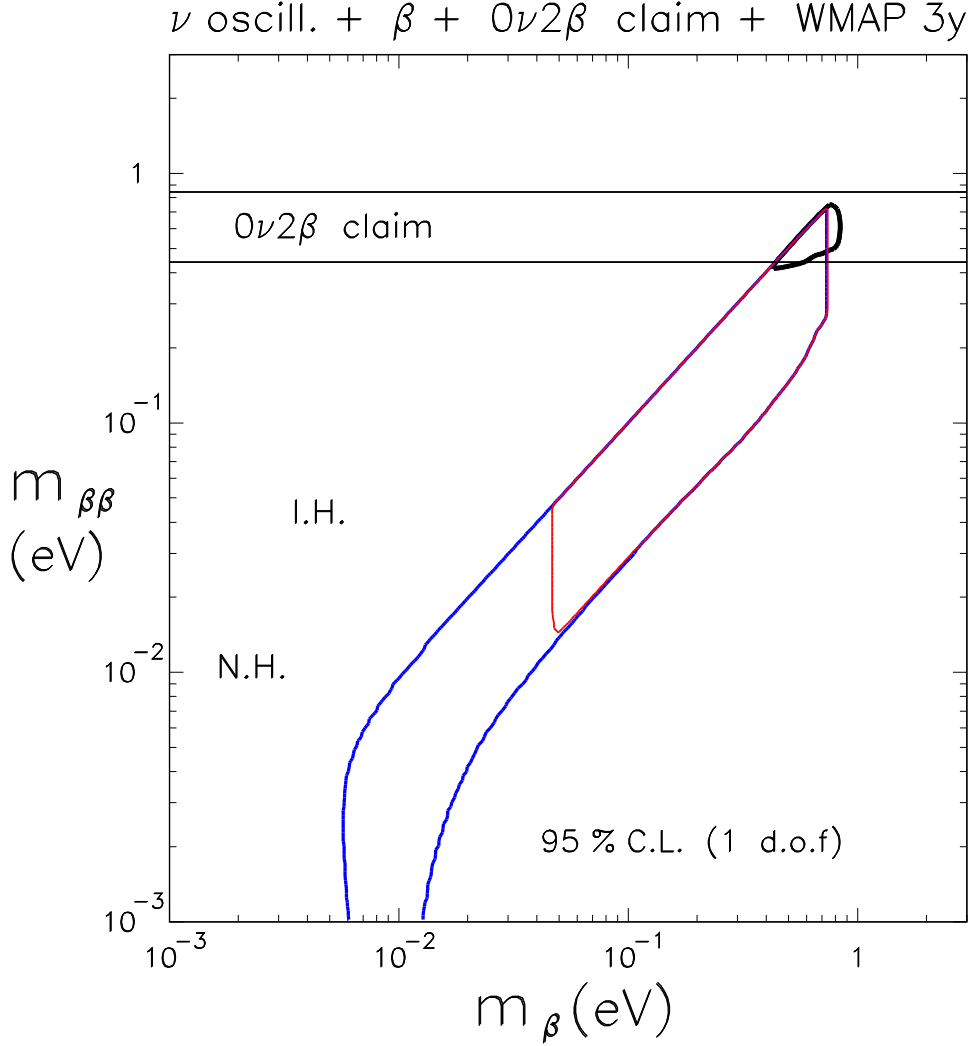}
\vspace*{2mm}
\caption{\label{fig6} 
As in Fig.~\protect\ref{fig5}, but in the plane
$(m_{\beta\beta},m_\beta$). }
\end{figure}

For the relatively high $\nu$ masses implied by the above $1^+$ 
case, the Mainz+Troitsk bound on $m_\beta$, included in the fit,
slightly tightens the global constraints
in the upper part of the wedge-shaped solution in Fig.~5
(this is the only such case). Not surprisingly,
if the global
combination dubbed ``1+'' were correct, a positive discovery
of $m_\beta$ should be ``around the corner,'' just below the
Mainz+Troitsk bound. For this reason, we find it useful
to represent in Fig.~6 the same case $1^+$, but in the plane $(m_{\beta\beta},m_\beta)$.
A  signal in the range  
 $m_{\beta}\sim\mathrm{few}\times 10^{-1}$ eV is clearly expected, and could be
 found in the next-generation
Karksruhe Tritium Neutrino Experiment (KATRIN) \cite{katrin}, which
should take data in the next decade.

\subsection{Case with all cosmological data but without the $0\nu2\beta$ claim}

\begin{figure}[t]
\vspace*{-0cm}\hspace*{-.5cm}
\includegraphics[scale=0.7]{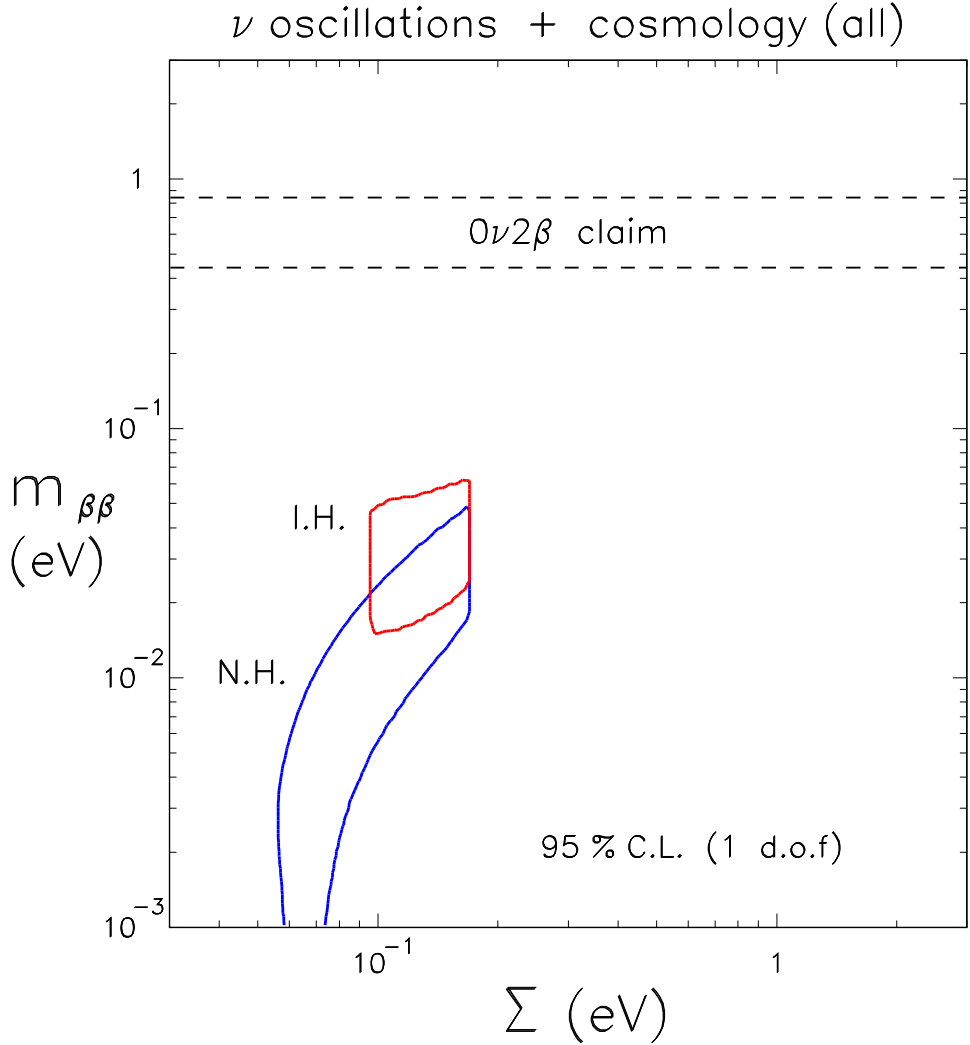}
\vspace*{2mm}
\caption{\label{fig7} 
Comparison (at $2\sigma$ level) between the regions allowed by 
the $0\nu2\beta$ signal claim (horizontal band) and by
neutrino oscillation data plus all cosmological data (slanted
bands) for normal (NH) and inverted (IH) hierarchy, in the
plane ($m_{\beta\beta},\Sigma$). In this case, $\beta$ decay data
constraints are not relevant, and oscillation plus cosmological data
cannot be combined with the $0\nu2\beta$ decay claim. This situation
corresponds to ``Case 7'' in the text.
}
\end{figure}

The most ``aggressive'' limit on $\Sigma$ is placed by all cosmological data
(case 7 in Table~\ref{tableCASES}),
assuming that their combination is correct and that there are no hidden systematics.
In this case, the upper limit on $\Sigma$ is so tight ($\sim 0.2$ eV), that no combination 
is possible with the $0\nu2\beta$ claim at face value. The incompatibility is
evident in Fig.~\ref{fig7}, which shows how far are the regions allowed by oscillation 
and cosmological neutrino data for both normal and inverted hierarchy 
(confined in the lower left corner) and the $0\nu2\beta$ claim
(horizontal band). However, one should not be tempted to say that 
``cosmological data rule out the $0\nu2\beta$ claim,'' for at least three reasons: 1)
such claim can only be (dis)proved by another $0\nu2\beta$ experiment 
with greater sensitivity; 2) the $0\nu2\beta$ signal might be due to new 
physics beyond light Majorana neutrinos; 3) independently of the correctness of 
the $0\nu2\beta$ claim, cosmological bounds on neutrino masses should be taken 
with a grain of salt, given the many systematic uncertainties that typically 
affect astrophysical data, and given our persistent ignorance about two main 
ingredients of the standard cosmological model (dark matter and dark energy).

\subsection{Overview of all cases}

From a glance at Fig.~\ref{fig4}, it appears that the $0\nu2\beta$ claim
is formally compatible (at $2\sigma$) with oscillation
data plus cosmological bounds only if the latter include just
WMAP-3y data (most conservative case 1). In this case, the global
combination dubbed as $1^+$ in Sec.~V~A becomes possible.
In all other cases, 
a combination would be formally possible only by stretching (some of) the errors
at (much) more than $2\sigma$ --- an attempt not pursued in this work.
Therefore, except for the case $1^+$ already examined, we consider
the possibility that the $0\nu2\beta$ claim is wrong (i.e., that it
provides an upper bound only), and combine oscillation data
with all cosmological datasets (1,...,7) in either normal and inverted hierarchy.

\begin{table}[t]
\caption{\label{tablePRED} Allowed ranges at 95\% C.L.\ ($2\sigma$)  
for the observables ($m_\beta,\, m_{\beta\beta},\, \Sigma$), as derived from
the global analysis of neutrino oscillation and non-oscillation data. 
The cases labeled from 1 to 7 (with either normal or inverted hierarchy)
include the seven cosmological input data sets discussed
in Sec.~IV, with the $0\nu2\beta$ decay
claim \cite{Kl04} treated as a one-sided (upper)
limit on $m_{\beta\beta}$. Only in the case $1^+$ we fully include the $0\nu2\beta$ decay claim 
in the global combination (the hierarchy being irrelevant in such case). 
See the text for details. }
\centering
\resizebox{0.87\textwidth}{!}{
\begin{ruledtabular}
\begin{tabular}{lcccc}
Case		&	Hierarchy	&	$m_\beta$ (eV)		& $m_{\beta\beta}$ (eV)	&	$\Sigma$ (eV)	\\[1mm]	
\hline
$1^+$	&	Any			&	$[0.42,\,0.84]$		&	$[0.42,\,0.75]$		&	$[1.3,\,2.5]$	\\[1mm]
\hline
1		&	Normal		&	$[0.0057,\,0.69]$	&	$[0,\,0.66]$			&	$[0.056,\,2.1]$	\\
2		&				&	$[0.0057,\,0.39]$	&	$[0,\,0.38]$			&	$[0.056,\,1.2]$	\\
3		&				&	$[0.0057,\,0.25]$	&	$[0,\,0.24]$			&	$[0.056,\,0.75]$	\\
4		&				&	$[0.0057,\,0.24]$	&	$[0,\,0.23]$			&	$[0.056,\,0.72]$	\\
5		&				&	$[0.0057,\,0.18]$	&	$[0,\,0.18]$			&	$[0.056,\,0.56]$	\\
6		&				&	$[0.0057,\,0.063]$	&	$[0,\,0.061]$		&	$[0.056,\,0.20]$	\\
7		&				&	$[0.0057,\,0.050]$	&	$[0,\,0.048]$		&	$[0.056,\,0.17]$	\\[1mm]
\hline
1		&	Inverted		&	$[0.047,\,0.69]$		&	$[0.015,\,0.66]$		&	$[0.095,\,2.1]$	\\
2		&				&	$[0.047,\,0.39]$		&	$[0.015,\,0.38]$		&	$[0.095,\,1.2]$	\\
3		&				&	$[0.047,\,0.25]$		&	$[0.015,\,0.24]$		&	$[0.095,\,0.75]$	\\
4		&				&	$[0.047,\,0.24]$		&	$[0.015,\,0.23]$		&	$[0.095,\,0.72]$	\\
5		&				&	$[0.047,\,0.18]$		&	$[0.015,\,0.18]$		&	$[0.095,\,0.56]$	\\
6		&				&	$[0.047,\,0.075]$	&	$[0.015,\,0.073]$	&	$[0.095,\,0.20]$	\\
7		&				&	$[0.047,\,0.064]$	&	$[0.015,\,0.062]$	&	$[0.095,\,0.17]$	\\[1mm]
\end{tabular}
\end{ruledtabular}
}
\end{table}

The results of such exercise 
are summarized in Table~\ref{tablePRED}, in terms of $2\sigma$
allowed ranges for the three relevant  observables $(m_\beta,m_{\beta\beta},\Sigma)$. 
Such ranges may be useful to gauge the sensitivity required
by future experiments, in order to explore the currently allowed parameter 
space. E.g., the $\beta$-decay experiment KATRIN, with an estimated sensitivity
down to $\sim 0.2$ eV \cite{katrin}, can fully test case $1^+$ in Table~\ref{tablePRED}, 
may test a fraction of the parameter space up to case 4, 
but it cannot access cases 5, 6, and 7, which will necessarily require new $\beta$-decay
 detection techniques to probe the $O(0.1$--$0.01)$ eV range for $m_\beta$. 
Analogously, in order to significantly
test the various cases in
Table~\ref{tablePRED},
cosmological data might need to reach a $2\sigma$ accuracy 
of about $\sim 0.1$ eV on $\Sigma$ (in the global fit), which may be
feasible in the future \cite{pastor}; on the other hand, 
$0\nu2\beta$ experiments might need to push their sensitivity down to
$O(0.1)$ eV or even to $O(0.01)$ eV, which is quite challenging \cite{Elli}. 
Cosmological data seems thus to be in a relatively good position 
to run the absolute neutrino mass race.


\section{Implications for future   $0\nu2\beta$ decay searches}

Concerning $0\nu2\beta$ experiments, we have shown in 
Table~\ref{tablePRED} the expected ranges for the ``canonical'' 
parameter $m_{\beta\beta}$ in various cases. However,  
in practice, 
the experimentalists measure or constrain decay half lives ($T^{0\nu}_{1/2}$)
rather than effective
Majorana masses ($m_{\beta\beta}$). Indeed, some confusion arises in the literature 
when the sensitivity of different $0\nu2\beta$ experiments is compared
in terms of $m_{\beta\beta}$, but the nuclear matrix
elements are not homogeneous, or they are not calculated within the same 
theoretical framework.
Therefore, we think it is useful to present also more ``practical''
predictions for 
half lives $T^{0\nu}_{1/2}$ in different nuclei, by using one and the same theoretical 
nuclear model \cite{faessler} with well-defined uncertainties 
(i.e., by using the matrix elements $C_{mm}$ as summarized in 
Table~\ref{tableNME}).

More precisely, 
the $m_{\beta\beta}$ ranges estimated in Table~\ref{tablePRED}, together with the
theoretical input values (and errors) for the nuclear matrix elements $C_{mm}$ reported in 
Table~\ref{tableNME}, are used to estimate the expected  $2\sigma$ ranges for the $0\nu2\beta$ 
half lives $T^{0\nu}_{1/2}$ in different nuclei
through Equation~(\ref{logs}). As previously remarked, 
the linear form of this equation makes it easier to propagate the uncertainties
affecting $C_{mm}$ and $m_{\beta\beta}$. As a result, we obtain 
in Table~\ref{tableNUCL} the $2\sigma$ ranges for the half-lives $T^{0\nu}_{1/2}$ of 
several candidate nuclei, for
all cases ($1^+,\, 1,\, 2,\dots,7$) previously considered in 
Sec.~V.

\begin{table*}[t]
\caption{\label{tableNUCL} Allowed ranges at 95\% C.L.\ ($2\sigma$)  
for the $0\nu2\beta$ decay half-lives $T^{0\nu}_{1/2}$ of various candidate
nuclei, for the same
cases as in Table~\protect\ref{tablePRED}. The $T^{0\nu}_{1/2}$  ranges are derived from the relation 
$T^{0\nu}_{1/2}=m^2_e/(m^2_{\beta\beta}\,C_{mm})$, by using the $C_{mm}$ and $m_{\beta\beta}$ input  
from Table~\protect\ref{tableNME} and Table~\protect\ref{tablePRED}, respectively.
The shorthand notation X.Y$e$ZW means $T^{0\nu}_{1/2}=\mathrm{X.Y}\times 10^\mathrm{ZW}$~yr.
For the sake of comparison, in the last row we list existing
experimental lower limits on $T^{0\nu}_{1/2}$ (at 90\% C.L.), 
as taken from the recent compilations in \cite{Elli,barabash}.}
\vspace*{4mm}
\centering
\resizebox{\textwidth}{!}{
\begin{tabular}{lcccccccc}
\hline
\hline
Case	& Hierarchy	& $^{76}\mathrm{Ge}$	&$^{82}\mathrm{Se}$	&$^{100}\mathrm{Mo}$	&$^{116}\mathrm{Cd}$	&$^{128}\mathrm{Te}$	&$^{130}\mathrm{Te}$	& $^{136}\mathrm{Xe}$ \\[4pt]
\hline
$1^+$& Any  		&$[1.0e25,\,3.5e25]$	&$[2.9e24,\,1.1e25]$	&$[5.4e24,\,2.3e25]$	&$[3.4e24,\,1.7e25]$	&$[7.2e25,\,4.2e26]$	&$[3.2e24,\,1.9e25]$	&$[5.2e24,\,8.7e25]$  \\[4pt]
\hline
1	& Normal		&$[1.1e25,\,\infty]$	&$[3.0e24,\,\infty]$	&$[5.1e24,\,\infty]$	&$[3.2e24,\,\infty]$	&$[6.5e25,\,\infty]$	&$[2.9e24,\,\infty]$	&$[4.3e24,\,\infty]$  \\
2	&   			&$[3.3e25,\,\infty]$	&$[8.9e24,\,\infty]$	&$[1.5e25,\,\infty]$	&$[9.5e24,\,\infty]$	&$[1.9e26,\,\infty]$	&$[8.7e24,\,\infty]$	&$[1.3e25,\,\infty]$  \\
3	& 			&$[8.3e25,\,\infty]$	&$[2.2e25,\,\infty]$	&$[3.9e25,\,\infty]$	&$[2.4e25,\,\infty]$	&$[4.9e26,\,\infty]$	&$[2.2e25,\,\infty]$	&$[3.2e25,\,\infty]$  \\
4	& 			&$[9.1e25,\,\infty]$	&$[2.5e25,\,\infty]$	&$[4.3e25,\,\infty]$	&$[2.6e25,\,\infty]$	&$[5.4e26,\,\infty]$	&$[2.4e25,\,\infty]$	&$[3.5e25,\,\infty]$  \\
5	& 			&$[1.4e26,\,\infty]$	&$[3.9e25,\,\infty]$	&$[6.8e25,\,\infty]$	&$[4.2e25,\,\infty]$	&$[8.5e26,\,\infty]$	&$[3.2e25,\,\infty]$	&$[5.6e25,\,\infty]$  \\
6	& 			&$[1.3e27,\,\infty]$	&$[3.4e26,\,\infty]$	&$[5.9e26,\,\infty]$	&$[3.6e26,\,\infty]$	&$[7.4e27,\,\infty]$	&$[3.3e26,\,\infty]$	&$[4.9e26,\,\infty]$  \\
7	& 			&$[2.1e27,\,\infty]$	&$[5.6e26,\,\infty]$	&$[9.8e26,\,\infty]$	&$[8.0e26,\,\infty]$	&$[1.2e28,\,\infty]$	&$[5.5e26,\,\infty]$	&$[6.5e26,\,\infty]$  \\ [4pt]
\hline
1	& Inverted	&$[1.1e25,\,3.3e28]$	&$[3.0e24,\,1.1e28]$	&$[5.1e24,\,2.5e28]$	&$[3.2e24,\,1.8e28]$	&$[6.5e25,\,4.7e29]$	&$[2.9e24,\,2.2e28]$	&$[4.3e25,\,1.1e29]$  \\
2	& 			&$[3.3e25,\,3.3e28]$	&$[8.9e24,\,1.1e28]$	&$[1.5e25,\,2.5e28]$	&$[9.5e24,\,1.8e28]$	&$[1.9e26,\,4.7e29]$	&$[8.7e24,\,2.2e28]$	&$[1.3e24,\,1.1e29]$  \\
3	& 			&$[8.3e25,\,3.3e28]$	&$[2.2e25,\,1.1e28]$	&$[3.9e25,\,2.5e28]$	&$[2.4e25,\,1.8e28]$	&$[4.9e26,\,4.7e29]$	&$[2.2e25,\,2.2e28]$	&$[3.2e25,\,1.1e29]$  \\
4	& 			&$[9.1e25,\,3.3e28]$	&$[2.5e25,\,1.1e28]$	&$[4.3e25,\,2.5e28]$	&$[2.6e25,\,1.8e28]$	&$[5.4e26,\,4.7e29]$	&$[2.4e25,\,2.2e28]$	&$[3.5e25,\,1.1e29]$  \\
5	& 			&$[1.4e26,\,3.3e28]$	&$[3.9e25,\,1.1e28]$	&$[6.8e25,\,2.5e28]$	&$[4.2e25,\,1.8e28]$	&$[8.5e26,\,4.7e29]$	&$[3.2e25,\,2.2e28]$	&$[5.6e25,\,1.1e29]$  \\
6	& 			&$[9.1e26,\,3.3e28]$	&$[2.5e26,\,1.1e28]$	&$[4.3e26,\,2.5e28]$	&$[2.6e26,\,1.8e28]$	&$[5.4e27,\,4.7e29]$	&$[2.4e26,\,2.2e28]$	&$[3.5e26,\,1.1e29]$  \\
7	& 			&$[1.2e27,\,3.3e28]$	&$[3.1e26,\,1.1e28]$	&$[5.4e26,\,2.5e28]$	&$[3.3e26,\,1.8e28]$	&$[6.8e27,\,4.7e29]$	&$[3.0e26,\,2.2e28]$	&$[4.5e26,\,1.1e29]$  \\ [4pt]
\hline
\multicolumn{2}{l}{Limits (90\% C.L.)}
				&$\gtrsim 1.2e25$ 			&$>2.1e23$ 			&$>5.8e23$ 			&$>1.7e23$ 			&$>7.7e24$ 			&$>2.2e24$ 			&$>4.5e23$ 			  \\[4pt]
\hline
\end{tabular}
}\vspace*{6mm}
\end{table*}

The last row in Table~\ref{tableNUCL} shows the current experimental limits on the various 
half lives, as taken from the recent compilations \cite{Elli,barabash}. It can be appreciated that, except for 
$^{76}$Ge and $^{130}$Te,  current limits 
must be improved by about one order of magnitude (at least) in order to probe any of the cases in 
Table~\ref{tableNME}. Even for the most promising nuclei 
($^{76}$Ge and $^{130}$Te), however, an experimental improvement by a factor of a few
in the $T^{0\nu}_{1/2}$ limits
can only allow to probe the most optimistic case $1^+$, and a fraction of 
the cases 1 and 2. Probing the other cases ($3,\dots,7$) 
 will require (much) more than 
an order-of-magnitude improvement with respect to current $0\nu2\beta$ limits. It is
thus very important to invest a great effort in future neutrinoless double beta decay
experiments with increasing sensitivity to longer lifetimes, in order to match the typical
expectations 
from current neutrino oscillation and cosmological data.


\section{Conclusions}

Building on previous work \cite{fogli} and on updated 
experimental input, 
we have revisited the phenomenological constraints applicable to
three observables sensitive to absolute neutrino masses: The
effective neutrino mass in single beta decay $(m_\beta)$; the
effective Majorana neutrino mass in neutrinoless double beta decay
$(m_{\beta\beta})$; and the sum of neutrino masses in cosmology
$(\Sigma)$. In particular, we have included the constraints coming from
the first MINOS results \cite{MINOS}
and from the WMAP-3y 
data \cite{WMAP3}, as well as other relevant cosmological data
and priors. We have found that the largest neutrino mass squared
difference $\Delta m^2$ is now
determined with a  15\% accuracy (at $2\sigma$).
We have also 
examined the upper bounds on the sum of neutrino masses $\Sigma$,
as well as their correlations with the
matter power spectrum normalization parameter $\sigma_8$. 
The bounds range  from 
$\Sigma \lesssim2$~eV (WMAP-3y data only) to 
$\Sigma\lesssim 0.2$~eV (all cosmological data)
at $2\sigma$, 
in agreement with previous studies. 
Finally, for various possible combination of data sets, 
we have revisited the (in)compatibility between current $\Sigma$ and 
$m_{\beta\beta}$ constraints, and have derived 
quantitative predictions for single and double beta-decay 
 observables, which can be useful to evaluate the 
sensitivity required by present and future experiments 
in order to explore the currently allowed parameter space.

\acknowledgments
The work of G.L.\ Fogli, E.\ Lisi, and A.\ Marrone is supported by the Italian
MIUR and INFN through the ``Astroparticle Physics'' research project. A.\ Palazzo
is supported by an INFN post-doc fellowship. A.\ Melchiorri
is supported by MIUR through the COFIN contract N.~2004027755.
We thank A.\ Bettini for useful remarks about an early version of the manuscript.

\end{document}